\documentclass[12pt]{article}

\begin{document}


\newcommand{\bb}{\begin{equation}}
\newcommand{\ee}{\end{equation}}
\newcommand{\bbb}{\begin{eqnarray}}
\newcommand{\eee}{\end{eqnarray}}
\newcommand{\diag}{\mbox{diag }}
\newcommand{\Str}{\mbox{STr }}
\newcommand{\Tr}{\mbox{Tr }}
\newcommand{\Det}{\mbox{Det }}
\newcommand{\C}[2]{{\lk [{#1},{#2}\re ]}}
\newcommand{\AC}[2]{{\lk \{{#1},{#2}\re \}}}
\newcommand{\kk}{\hspace{.5em}}
\newcommand{\vc}[1]{\mbox{$\vec{{\bf #1}}$}}
\newcommand{\mc}[1]{\mathcal{#1}}
\newcommand{\del}{\partial}
\newcommand{\lk}{\left}
\newcommand{\ave}[1]{\mbox{$\langle{#1}\rangle$}}
\newcommand{\re}{\right}
\newcommand{\pd}[1]{\frac{\del}{\del #1}}
\newcommand{\pdd}[2]{\frac{\del^2}{\del #1 \del #2}}
\newcommand{\Dd}[1]{\frac{d}{d #1}}
\newcommand{\sech}{\mbox{sech}}
\newcommand{\pref}[1]{(\ref{#1})}

\newcommand
{\sect}[1]{\vspace{20pt}{\LARGE}\noindent
{\bf #1:}}
\newcommand
{\subsect}[1]{\vspace{20pt}\hspace*{10pt}{\Large{$\bullet$}}\mbox{ }
{\bf #1}}
\newcommand
{\subsubsect}[1]{\hspace*{20pt}{\large{$\bullet$}}\mbox{ }
{\bf #1}}

\def\ie{{\it i.e.}}
\def\eg{{\it e.g.}}
\def\cf{{\it c.f.}}
\def\etal{{\it et.al.}}
\def\etc{{\it etc.}}

\def\AA{{\cal A}}
\def\BB{{\cal B}}
\def\CC{{\cal C}}
\def\DD{{\cal D}}
\def\EE{{\cal E}}
\def\FF{{\cal F}}
\def\GG{{\cal G}}
\def\HH{{\cal H}}
\def\II{{\cal I}}
\def\JJ{{\cal J}}
\def\KK{{\cal K}}
\def\LL{{\cal L}}
\def\MM{{\cal M}}
\def\NN{{\cal N}}
\def\OO{{\cal O}}
\def\PP{{\cal P}}
\def\QQ{{\cal Q}}
\def\RR{{\cal R}}
\def\SS{{\cal S}}
\def\TT{{\cal T}}
\def\UU{{\cal U}}
\def\VV{{\cal V}}
\def\WW{{\cal W}}
\def\XX{{\cal X}}
\def\YY{{\cal Y}}
\def\ZZ{{\cal Z}}

\def\sinh{{\rm sinh}}
\def\cosh{{\rm cosh}}
\def\tanh{{\rm tanh}}
\def\sgn{{\rm sgn}}
\def\det{{\rm det}}
\def\trace{{\rm Tr}}
\def\exp{{\rm exp}}
\def\sh{{\rm sh}}
\def\ch{{\rm ch}}

\def\ell{{\it l}}
\def\str{{\it str}}
\def\lp{\ell_{{\rm pl}}}
\def\blp{\overline{\ell}_{{\rm pl}}}
\def\ls{\ell_{{\str}}}
\def\bls{{\bar\ell}_{{\str}}}
\def\bM{{\overline{\rm M}}}
\def\gs{g_\str}
\def\gym{{g_{Y}}}
\def\geff{g_{\rm eff}}
\def\eff{{\rm eff}}
\def\r11{R_{11}}
\def\kel{{2\kappa_{11}^2}}
\def\kten{{2\kappa_{10}^2}}
\def\lpten{{\lp^{(10)}}}
\def\alp{{\alpha '}}
\def\alpe{{{\alpha_e}}}
\def\le{{{l}_e}}
\def\aleff{{\alp_{eff}}}
\def\sqaleff{{\alp_{eff}^2}}
\def\tgs{{\tilde{g}_s}}
\def\talp{{{\tilde{\alpha}}'}}
\def\tlp{{\tilde{\ell}_{{\rm pl}}}}
\def\tr11{{\tilde{R}_{11}}}
\def\wtilde{\widetilde}
\def\what{\widehat}
\def\hlp{{\hat{\ell}_{{\rm pl}}}}
\def\hr11{{\hat{R}_{11}}}
\def\hf{{\textstyle\frac12}}
\def\coeff#1#2{{\textstyle{#1\over#2}}}
\def\CY{Calabi-Yau}
\def\lessapprox{\;{\buildrel{<}\over{\scriptstyle\sim}}\;}
\def\greaterapprox{\;{\buildrel{>}\over{\scriptstyle\sim}}\;}
\def\inbar{\,\vrule height1.5ex width.4pt depth0pt}
\def\IC{\relax\hbox{$\inbar\kern-.3em{\rm C}$}}
\def\IR{\relax{\rm I\kern-.18em R}}
\def\IP{\relax{\rm I\kern-.18em P}}
\def\Z{{\bf Z}}
\def\R{{\bf R}}
\def\One{{1\hskip -3pt {\rm l}}}
\def\sst{\scriptscriptstyle}
\def\osc{{\rm\sst osc}}
\def\lam{\lambda}
\def\lc{{\sst LC}}
\def\pr{{\sst \rm pr}}
\def\cl{{\sst \rm cl}}
\def\D{{\sst D}}
\def\bh{{\sst BH}}
\def\vev#1{\langle#1\rangle}

\input{epsf}

\begin{titlepage}
\rightline{CLNS 00/1699}

\rightline{hep-th/0010237}

\vskip 2cm
\begin{center}
\Large{{\bf 
Transcribing spacetime data\\ into matrices
}}
\end{center}

\vskip 2cm
\begin{center}
Vatche Sahakian\footnote{\texttt{vvs@mail.lns.cornell.edu}}
\end{center}
\vskip 12pt
\centerline{\sl Laboratory of Nuclear Studies}
\centerline{\sl Cornell University}
\centerline{\sl Ithaca, NY 14853, USA}

\vskip 2cm

\begin{abstract}

In certain 
supergravity backgrounds,
D0 branes may polarize
into higher dimensional Dp branes. We study this phenomenon 
in some generality
from the perspective of a local inertial observer and explore 
polarization effects resulting from tidal-like forces. We find
D2 brane droplets made of D0 branes at an extremum of the Born-Infeld
action even in scenarios where the RR fields may be zero. These solutions lead us to
a local formulation of
the UV-IR correspondence. A holographic Planck scale bound on the number of
D0 branes plays an important role in the analysis.
We focus on the impact of higher order
moments of background fields
and work out extensions of the non-commutative algebra
beyond the Lie and Heisenberg structures. In this context, it appears that 
q-deformed algebras come into play.

\end{abstract}

\end{titlepage}
\newpage
\setcounter{page}{1}

\section{Introduction}
\label{intro}

Recently, a deeper understanding
of properties of D branes
in supergravity backgrounds was 
achieved in the work of~\cite{MEYERS,WATIDBI}. A host of new interaction terms 
were identified in the Dirac-Born-Infeld (DBI) action that explore
the non-commutative dynamics of the D-brane coordinates.
In particular, it was pointed out
that RR fields carrying the charges of Dp branes, along with momentum modes in
certain special scenarios, can polarize
D0 branes~\cite{MEYERS,GIANT,HASHHIRITZ,DASGIANT,DTVPOL}. 
The resulting objects appear
to be microscopic realizations 
of higher dimensional Dp branes. Such configurations
have typically non-commutative field theories living on their 
worldvolumes~\cite{SWNC,AOKI,SEIBBACK}, 
and encode in their shape information about the
background space.
D-branes are natural probes of the Planck scale 
structure of space-time~\cite{DKPS}.
They entail exotic dynamics, involving stringy non-local interactions,
and are non-perturbative in character~\cite{POLCHV1}. 
It then becomes important to understand how data in
the spacetime fields gets transcribed through polarization
into the matrices representing the coordinates of D branes.

In this work, we investigate couplings of D0 branes to most of the
supergravity bosonic fields of the IIA theory from the perspective of
a local inertial observer. In addition to polarization effects
from RR gauge fields, tidal-like forces arising from various interaction
terms in the DBI action
can also polarize the D0 branes.
We find ellipsoidal droplets of D0 branes that store
some of the information about
the supergravity background in their shape.
Furthermore, in certain
regions of the background field parameter space where we may
naively expect polarization by tidal forces, these ellipsoids are unstable. 
There is a game of competition between couplings of the D branes to the
supergravity fields and the forces binding the $N$ branes together
through strings stretched between them. By looking for such
configurations at the extrema of the DBI action, we effectively
probe into the distribution
of these forces within the polarized ellipsoid.

Throughout our discussion, we will be ignoring
effects of back-reaction from the D0 branes onto the background spacetime.
Within this approximation scheme,
we find that the regime of validity of the expansion of the DBI action
arises as a statement bounding the number of D0 branes by a local
measure of area in Planck units. Given a characteristic
length scale $L$ for background field
variations (relating to, for example, the local scale for the
spacetime curvature), $N$ D0 branes living in a three dimensional
subset of the transverse space must satisfy the bound $N\ll L^2/\lp^2$. 
A second observation is
a local realization of the UV-IR correspondence~\cite{PEETPOLCH}. 
We argue that the 
scale of non-commutativity in the worldvolume theory of the polarized
D0 brane configuration is inversely proportional to the length scale
characterizing local variations in the background fields.

To store information about 
higher moments of
background fields into D0 brane configurations, one needs to
go beyond the Lie algebraic structure. 
We present a prescription on how to encode this additional data in
generalized algebras. We find that, at the next order in the expansion,
the spacetime gets transcribed into
algebras with q-deformed structure~\cite{QWESS,MSSW}.

We note that the phenomenon we investigate has to do with
polarization effects involving a large number of D0 branes. Recently, other
authors have explored somewhat related dynamics arising
from the fermionic degrees of freedom 
on a single D0 brane~\cite{MILLARWATI,HO}.

In Section~\ref{mainsec}, we setup the actions and equations of motion
of interest; we present several simple solutions and analyze the
underlying dynamics.  In Section~\ref{qgsec}, we consider effects
higher order in the string tension, and describe extensions of Lie algebras
that solve the equations of motion.
In Section~\ref{ncsec}, we briefly outline a non-compact solution
describing a non-commutative hyperboloid that can be realized with
infinite size matrices. The Appendix contains a few technical details
used in the main text.

{\bf Note added: }
The solutions we find are at an extremum of the energy and there is an issue
of stability that needs to be addressed.
I am grateful to M. Spradlin and
A. Volovich for bringing this issue to my attention. We comment on this problem
briefly in the text, and defer a detailed analysis to an upcoming
work~\cite{STABLE}.

\section{Polarization with Lie algebraic structure}
\label{mainsec}

In this section, we study static solutions with U(2) algebraic
structure describing N D0 branes in background supergravity fields.
In the first subsection, we set up the action and
the equations of motion to cubic order in the inverse string tension. 
In Section~\ref{ellipsoidsec},
we write solutions for backgrounds where all fields
but the D2 brane gauge field are nonzero. In Section~\ref{analsec},
we address some issues regarding the stability of these
configurations and the regime of validity of our classical calculation,
formulating a local statement regarding 
the UV-IR correspondence.
We end with Section~\ref{balancesec} by considering the effect of the D2 brane
gauge field on the dynamics. Some details of this section
are sketched in the Appendix.

\subsection{The setup and equations of motion}
\label{setupsec}

Consider
$N$ D0 branes emersed in a general type IIA supergravity background. 
The branes are described by $N\times N$ hermitian
matrices $\Phi^i$, with $i=1..9$.
The dynamics in the energy regime 
of interest is governed by the non-Abelian DBI action~\cite{MEYERS}
\bbb\label{maineq}
S&=&-\frac{1}{\gs\ls} \int dt\ \Str \lk \{e^{-\phi}
\lk( -\lk(P\lk [E_{00}+E_{0i} \lk( Q^{-1} - \delta \re)^{ij} E_{j0}\re]
\re)\re)^{1/2} \lk( \det Q\re)^{1/2}\re \} \nonumber \\
&+& \frac{1}{\gs\ls} \int \Str \lk\{ P\lk [ e^{i\lambda i_\Phi i_\Phi } 
\lk( \sum C^{(n)} e^B\re) \re ] \re\}\ ,
\eee
where the matrix $Q$ is defined by
\bb
Q^i_j\equiv \delta^i_j + i \lambda \C{\Phi^i}{\Phi^k} E_{kj}\ ,
\ee
and
\bb
E_{\mu\nu}\equiv G_{\mu\nu} + B_{\mu\nu}\ .
\ee
We will ignore the back reaction effects of the D0 branes on the background.
Throughout, we will follow closely the notation introduced in~\cite{MEYERS}.
The space-time metric and NS-NS gauge field are denoted by $G_{\mu\nu}$ 
and $B_{\mu\nu}$ respectively; we also have the dilaton field $\phi$ and
the RR gauge fields $C^{(n)}$;
$\lambda\equiv 2\pi \ls^2$ is the inverse string tension, 
and $\gs$ is the IIA string coupling.
The $\Phi^i$'s appear also in equation~\pref{maineq} implicitly
through the dependence of the supergravity fields on the spacetime
coordinates~\cite{GARMEY1,GARMEY2,MEYERS}
\bb\label{expansion}
\psi\equiv \lk.e^{\lambda \Phi^i \del_i} \psi(x)\re|_P\ ,
\ee
where $P$ is a point about which we expand the fields, and 
$\psi$ represents any of the supergravity fields;
this corresponds to a standard normal ordering prescription. $\Phi^i$'s
are also hidden in the
pull-back of the fields to the world volume of the D0 branes, denoted by 
$P[\psi_0]\equiv \psi_0 +\lambda \dot{\Phi}^i \psi_i$ in the
canonical static gauge, where the dot denotes differentiation with respect
to time. Finally, $i_\Phi$ denotes the interior product operator, and 
$\Str$ is short-hand for the symmetrized trace prescription first introduced
in~\cite{TSEYTDBI}.

In the cases of interest, it will be useful to rescale the metric so as
to eliminate the dilaton field appearing in front of the DBI action. 
This trick will allow us to consider 
geodesic motion that incorporates the effects of the coupling of the D0
branes to the dilaton field~\footnote{In a previous version of this paper,
this approach was not used and the effect of the dilaton field appeared
explicitely in subsequent equations. The current approach is clearer and
simplifies the discussion.}.
Consider the scenario where we let go of a number of D0 branes in
a generic background field configuration, and observe the center of mass of the D0 branes as it
follows a geodesic in the rescaled metric.
The viewpoint of the freely falling D0 branes is captured by choosing
Fermi normal coordinates~\cite{MISNER}
\bb\label{inertial}
\lk.G_{\mu\nu}\re|_P=\eta_{\mu\nu}\ \ \ ,
\lk.G_{\mu\nu,\alpha}\re|_P=0\ ,
\ee
where $P$ is a point along the geodesic. It is known that the
metric seen in this reference frame will vary in time adiabatically,
so that one can consider static solutions at an instant, and evolve them
in time trivially. We also assume that all other supergravity fields vary 
slowly in the observer's time coordinate as well. This can sometimes be arranged
by a judicious choice of the initial conditions for the geodesic motion.
We then drop all time derivatives of all fields, and look for static
configurations of $N$ D0 branes in this freely falling frame with
\bb
\dot{\Phi}^i=0\ .
\ee
Using constant gauge
transformations, we can set the value of any gauge field at point $P$
equal to any constant value; for example, we choose
\bb
\lk.B_{\mu\nu}\re|_P=0\ .
\ee
Furthermore, the coupling $\gs$ in front of~\pref{maineq} is taken as the value of
$e^\phi$ at point P and is not used in the rescaling of the metric.
We then expand~\pref{maineq} in powers of $\lambda$. 
We will
quantify the regime of validity of this expansion more carefully later.
As a statement relevant only to terms of order $\lambda^3$ and beyond, 
we require additionally that
\footnote{Part of the motivation for this statement is that the 
effect of the NS-NS electric field
on D branes may be better explored with D-instantons probes.
More on this issue in the Discussion section.}
\bb
B_{ti,j}=0\ .
\ee
The main technical 
simplifications that results in this game
are that the term $(Q^{-1}-\delta)$ in~\pref{maineq}
does not contribute to order $\lambda^3$, and that the pull-back is trivial.

The strategy is to expand the DBI action while dropping all terms involving
only matrices proportional to the identity. This is because these terms describe
the dynamical evolution of the center of mass of the D0 branes, which we will solve for
separately using the full DBI action without an expansion. Stated differently, 
the terms proportional to the identity will be summed back to the square root and
the U(1) part of the ansatz for the $\Phi^i$'s will involve time dependence. 
The rest of the problem
involves looking for static solutions in SU(N), with the U(1) factored out.
This procedure assumes that the size of the D0 branes in the center of mass frame
will not affect its center of mass trajectory. This would be the case if
the size of the polarized D0 branes is much smaller than the typical wavelength
over which the background fields vary; so that the center of mass dynamics is that
of a point-like object. To order $\lambda^2$, this decoupling of the U(1) and SU(N)
sectors in the dynamics can be easily seen. To order $\lambda^3$, this is more 
subtle~\footnote{In a previous version of this work, this decoupling issue was
addressed explicitely, culminating in finding (redundantly) unstable modes
in the U(1) sector. In our current approach, this issue gets circumvented and
the treatment is more transparent.}.

Consider first an expansion to order $\lambda^2$; the form of the action can be
determined (almost uniquely) by noting that each $\Phi^i$ arising
in a symmetric combination and each commutator of the $\Phi^i$'s
come with a power of $\lambda$, and by making use of the
symmetrized trace prescription.
We are led to the structure\footnote{
We note that 
the $\Phi^i$'s have dimension of inverse length; \ie\ the
combination $\lambda \Phi^i$ relates to spacetime coordinates.}
\bbb\label{actiontwo}
S&=&-\frac{1}{\gs\ls}\int dt\ \Str\lk \{ 
+ \frac{\lambda^2}{2} M_{ij} \AC{\Phi^i}{\Phi^j}\re. \nonumber \\
&+&\lk. i \lambda^2 N_{ikl} \C{\Phi^i}{\Phi^k} \Phi^l
+\lambda^2 P_{ijkl} \C{\Phi^l}{\Phi^k}\C{\Phi^j}{\Phi^i} +O(\lambda^3)\re\}\ ,
\eee
where $M_{ij}$, $N_{ijk}$ and $P_{ijkl}$ are c-number background
fields evaluated at point $P$.
These fields then acquire the following properties
\bb
M_{ij}=M_{ji}\ \ \ ,\ \ \ 
N_{ijk}=N_{[ijk]}\ ;
\ee
\bb\label{prel}
P_{ijkl}=-P_{jikl}\ \ \ ,\ \ \ 
P_{ijkl}=-P_{ijlk}\ \ \ ,\ \ \ 
P_{[ijk]l}=0\ \ \Rightarrow\ \  
P_{ijkl}=P_{klij}\ .
\ee
Hence, the $P$ field has the properties of a Riemann curvature
tensor; not surprisingly, since the commutator structure multiplying
it in the action
is related to a curvature form on a Lie algebra defined by the $\Phi^i$'s. 

Expanding~\pref{maineq}, we obtain the following realizations for the
background fields 
\bbb
M_{ij}&\equiv &  -\frac{1}{4} G_{00,ij}-\frac{1}{2} C^{(1)}_{0,ij} \label{mij}
\ ;\\
N_{ikl}&\equiv & \frac{1}{2} B_{[ki,l]}-\frac{1}{2} C^{(3)}_{0[ki,l]}
-\frac{1}{2} C^{(1)}_0 B_{[ki,l]}
\ ;\\
P_{ijkl}&\equiv & \frac{1}{4} \delta_{kj}\delta_{li} +
\frac{1}{8} C^{(5)}_{ijkl0} \rightarrow \frac{1}{4} \delta_{kj}\delta_{li}
\label{Peq}\ .
\eee
It is important to emphasize that all supergravity fields appearing
in these equations are evaluated at the point $P$\ \footnote{
Note that the value of the dilaton at point $P$ is factored out in
front of the action and the rest of the dilaton field was absorbed in the 
metric, denoted here as $G_{\mu\nu}$.}. We also have made
use of the choices for coordinates and gauges described above.
In the last equation, the D4 brane gauge field does not
contribute due to the symmetry relations~\pref{prel}\footnote{It is amusing to 
note that, in the three dimensional setting we will be focusing on,
the ``Weyl tensor'' associated with the $P$ field vanishes; hence, the
content of this field in general
is that of a symmetric two tensor. Yet, we have checked that, in looking
for solutions to our equations, the only ``physically non-trivial''
content of $P$ comes about from the term shown in~\pref{Peq}.}.

The equations of motion that follow from~\pref{actiontwo} are
\bb\label{eomtwo}
2 \lambda M_{in} \Phi^i
+3 i\lambda N_{nkl}\C{\Phi^k}{\Phi^l}
+\lambda \C{\Phi^j}{\C{\Phi^j}{\Phi^n}}=0\ .
\ee

In a second scenario of interest, we would like to study the
effect of the couplings
appearing in~\pref{maineq} at order $\lambda^3$. To 
avoid overwhelming ourselves with too much new physics, we will set all
fields to zero, except the metric and the dilaton. Expanding~\pref{maineq}
to order $\lambda^3$, we get
\bbb\label{actionthree}
S&=&-\frac{1}{\gs\ls}\int dt\ \Str\lk \{ \frac{\lambda^2}{2} M_{ij} \AC{\Phi^i}{\Phi^j}
+\frac{\lambda^2}{4} \C{\Phi^i}{\Phi^j}\C{\Phi^j}{\Phi^i} \re. \nonumber \\
&+& \lk. \lambda^3 \Phi^i \Phi^j \Phi^k T_{ijk}+O(\lambda^4)\re\}\ .
\eee
We have defined
\bb
T_{ijk}\equiv -\frac{1}{12} G_{00,ijk}\ .
\ee
The equations of motion become
\bbb\label{eomthree}
2 \lambda M_{in} \Phi^i
+\lambda \C{\Phi^j}{\C{\Phi^j}{\Phi^n}}
+\frac{3}{2} \lambda^2 \AC{\Phi^i}{\Phi^j} T_{nij}=0\ .
\eee

Looking for solutions to~\pref{eomtwo} and~\pref{eomthree},
we focus on backgrounds which are non-trivial only in 
a three dimensional subset of the full nine dimensional space; \ie\ there are
no background fields turned on with components in the other six space 
dimensions, and all fields have only dependence on coordinates within the chosen
three dimensional subset. This means that the indices $i,j,k,...$
run from one to three. We then have three matrices $\Phi^i$ appearing in
the action for which we will need to solve for, while the rest are set to zero.
Physically, this means that we have arranged for a situation where the
D0 branes sit on top of each other in six of the nine
dimensions of the transverse space, and acquire non-trivial configurations
in the remaining three.
Later on, we will briefly comment on generalizations of this
setup to situations that explore larger dimensional background spaces.

In Section~\ref{ellipsoidsec}, we will look for solutions to~\pref{eomtwo}
with Lie algebraic structure. The relevant algebra
is SU(2), and deformations and contractions
of it. The matrices are in an $N\times N$
representation. The case where $M_{ij}=0$
was discussed in~\cite{MEYERS}. To order $\lambda^2$, new
physics will arise from the term $M_{ij}$, which encodes, partly,
the effects of polarizing the $N$ D0
branes by the background curvature of space, \ie\ by the 
tidal forces in the local inertial frame.

In Section~\ref{qgsec},
we will study solutions of~\pref{eomthree} and  find
that Lie algebraic structure is insufficient to encode
all of the data in the background fields into matrices; the algebra that
solves these equations will acquire structure similar to ones that
arise in the context of q-deformed algebras.

\subsection{The non-commutative ellipsoid}
\label{ellipsoidsec}

We consider solutions of~\pref{eomtwo} in backgrounds
confined to a three dimensional subset of the full nine dimensional
transverse space, as described in the previous section.  
The field $N_{ijk}$ is then given by
\bb
N_{ijk}=\tilde{N} \varepsilon_{ijk}\ ,
\ee
where
\bb
\tilde{N}=\frac{1}{2} \lk( C^{(3)}_{012,3} +H_{123} \lk(C^{(1)}_0-1\re)\re)\ .
\ee
We look for configurations of matrices $\Phi^i$ in SU(2)
\bb
\Phi^i=\sigma^i\ ,
\ee
that solve~\pref{eomtwo}, with the $\sigma^i$'s obeying
\bb
\C{\sigma^i}{\sigma^j}=C^{ij}_{\kk k} \sigma^k\ .
\ee
The unknowns are the structure constants
and we write them as
\footnote{
The $\varepsilon$ tensor is written in flat background; \ie\ we accord
no significance to the location of the indices on it.
}
\bb
C^{ij}_{\kk k}=i\ \varepsilon^{ijl}\ g_{lk}\ ,
\ee
where the metric $g_{lk}$ is proportional to the Cartan-Killing metric
\footnote{
Without loss of generality, we can assume that this constant of proportionality
is positive, \ie\ the metric $g_{ij}$
is positive definite; it is also non-singular, and symmetric.
Given that we are mapping spacetime indices to indices in group space,
this corresponds to our freedom to align the orientations of the two frames
with respect to each other.
}.
Equation~\pref{eomtwo} determines this metric;
\ie\ we are looking for the particular linear combination of the canonical
SU(2) matrices that solve the equation. This leads us to
\bb\label{mateq}
2 M_{ln}- 6 \tilde{N} g_{nl}  -g_{lr} g_{rn} + g_{ii} g_{nl}=0\ .
\ee

Let us first look at the case where $M_{ij}=0$. Then equation~\pref{mateq} is
linear in $g_{ij}$ and the solution is simply
\bb
C^{kn}_{\kk m}= 3 i \tilde{N} \varepsilon^{knm}\ .
\ee
This was the case considered in~\cite{MEYERS}. 
The configuration is a non-commutative two sphere
describing D0 branes polarized by the $\tilde{N}$ field.

\begin{figure}
\epsfysize=5.5cm \centerline{\leavevmode \epsfbox{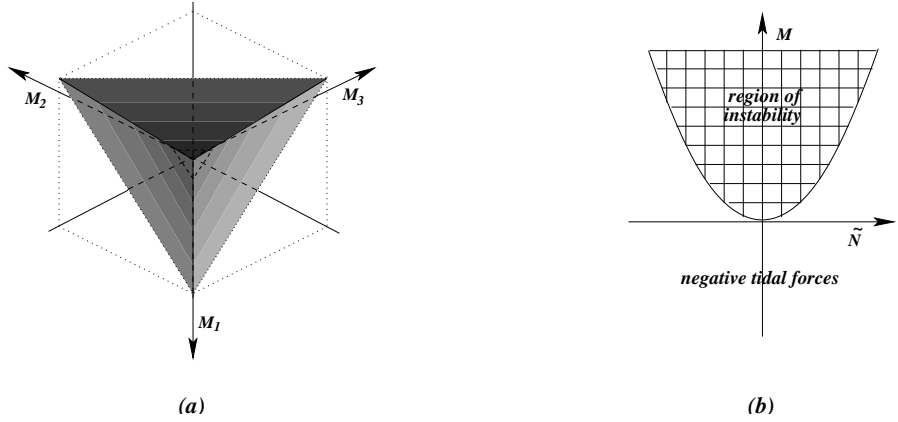}}
\caption{\sl (a) The parameter space of the non-commutative
ellipsoid. The region within the pyramid is associated with polarized states.
The apex of the pyramid is at the origin and
the region depicted is the negative quadrant;
(b) The phase structure of the D0 brane vacuum configurations in the
presence of an isotropic $M_{ij}$ and $\tilde{N}$ background fields. Everywhere
outside the shaded region, the $g_+$ solution of~\pref{gpm} prevails. 
}
\label{fig1}
\end{figure}

Next, consider the case where
$\tilde{N}=0$ and $M_{ij}$ is non-trivial. Equation~\pref{mateq}
is then a simple quadratic matrix equation.
The coordinate system we have chosen, the inertial frame at point $P$,
leaves a remnant of spacetime coordinate invariance at our disposal.
We can use an SO(3) subset of this gauge
freedom to diagonalize $M_{ij}$. We write
$M_{ij}=\diag(M_1,M_2,M_3)$ and define
\bb
a_1\equiv \lk(M_1-M_2-M_3\re)^{1/2}\ \ ,\ \ 
a_2\equiv \lk(M_2-M_1-M_3\re)^{1/2}\ \ ,\ \ 
a_3\equiv \lk(M_3-M_2-M_1\re)^{1/2}\ .
\ee
The only symmetry of equation~\pref{mateq} is the group $S_3$ permuting
the eigenvalues of $M_{ij}$. In this coordinate system, the
only solution is then given by $g_{ij}=\diag(g_1,g_2,g_3)$ with
\bb
g_{1}=+\frac{a_2 a_3}{a_1}\ \ ,\ \ g_{2}=+\frac{a_1 a_3}{a_2}\ \  ,\ \  
g_{3}=+\frac{a_2 a_1}{a_3}\ .
\ee
We also have the solution with negative $g_i$'s, corresponding to 
the physically equivalent situation $\Phi^i\rightarrow -\Phi^i$ (see
footnote 9 above).  We need to require 
\bb\label{pyramid}
a_i^2 \geq 0\ \ \ \forall\ i\ .
\ee
Otherwise, the Cartan-Killing metric is complex, and the matrices $\Phi^i$
must be {\em anti-}hermitian; \ie\ the coordinates of the D0 branes would be
complex. We may consider the case where any two of the $a_i^2$'s 
are negative; however, this brings one of the metric eigenvalues 
onto the other branch of the square root; the signature of the Cartan
metric changes and the corresponding group is
not compact; it cannot be embedded in a compact SU(N). 
The parameter space available to us by condition~\pref{pyramid} is depicted in
Figure~\ref{fig1}(a). 
The allowed region is where all the eigenvalues of $M_{ij}$
are non-positive and lie within the shaded pyramid. 
The normalization between the canonical form of the SU(2)
generators $\C{\tau^i}{\tau^j}=2\ i\ \varepsilon_{ijk} \tau^k$ and ours is 
\bb
\Phi^i=\frac{a_i}{2\sqrt{2}}\tau^i\ .
\ee
The D0 branes form a non-commutative ellipsoid encoding the content
of the background field $M_{ij}$ in the sizes of the radii.
The sides of the pyramid in Figure~\ref{fig1}(a) are singular planes where one
of the $a_i$'s vanishes and the ellipsoid gets squashed into a disk. 
We have a singular solution
also when one of the eigenvalues of $M_{ij}$ is zero; then the other two must be
equal and two of the $a_i$'s vanish; we are then at one of the edges
of the pyramid. The ellipsoid has collapsed in this scenario into an 
infinitely thin cigar. 
The solution fails when two or more eigenvalues of the
background $M_{ij}$ matrix vanish. In backgrounds that explore
the parameter space beyond the shaded pyramid, 
we need to look for structurally
different solutions; we will
come back to this issue in Section~\ref{ncsec}.

We also should look at the potential energy of this configuration,
to ascertain that these are energetically favored over the trivial configuration.
We will need the trace of the $\sigma^i$
matrices, which can be easily found using the Wigner-Eckart theorem
\footnote{
The Cartan-Killing metric is given by
\bb\label{cartan}
k^{ij}= 2\ a_1 a_2 a_3\ g^{ij}=C^{ik}_{\kk l} C^{jl}_{\kk k}\ .
\ee
}
\bb\label{hij}
h^{ij}\equiv 
\Tr \lk\{ \sigma^i \sigma^j\re\} = a_i a_j \frac{(N^2-1)N}{12} \delta^{ij}
\mbox{\kk\kk\kk\kk\kk\kk (no sum over }i,j\mbox{)}\ .
\ee
We write the potential energy in the general case when $\tilde{N}\neq 0$
in a suggestive form
\bb
V=\frac{1}{\gs \ls} \lambda^2 h^{ij}
\lk(\tilde{N} g_{ij} + g_{ik}g_{kj}-\frac{1}{2} g_{kk} g_{ij}\re)\ ,
\ee
where we have used equation~\pref{mateq}.
For the case at hand, with $\tilde{N}=0$ and $h^{ij}$ given by~\pref{hij}, 
we get
\bb\label{potone}
V=-\frac{1}{\gs \ls} \frac{\lambda^2}{24} N (N^2-1)\  ( a_1 a_2 a_3 )\ g_{ii}\ .
\ee
The potential energy for the configuration where all the D0 branes
sit on top of each other at $\Phi^i=0$ is zero. To claim a preferred
or competing configuration, we need that the energy given by~\pref{potone}
be non-positive.  The $M_i$'s are non-positive, while
the metric $g_{ij}$ and the $a_i$'s are all positive definite. We then always
have $V< 0$. There are also configurations where all of the $\Phi^i$'s are in
the Cartan subalgebra of SU(N). These correspond to non-static scenarios 
that can boundlessly lower the energy. We will address their relevance 
in the dynamics and to the stability of our solutions in~\cite{STABLE}.

\subsection{Analyzing the ellipsoid}
\label{analsec}

Physically, in the region of the parameter space where the various
conditions outlined above are satisfied, we are describing 
a puffed ellipsoid of D0 branes, with the 
relative sizes of the three radii of the ellipsoid being
related to the anisotropy in $M_{ij}$. 
Evaluating the components of the Riemann tensor in the local
inertial coordinate system at point $P$, we have
\bb
R_{i0j}^{\kk\kk 0}=\frac{1}{2} G_{00,ij}=R_{ij}-R_{ikj}^{\kk\kk k}
\equiv \tau_{ij}-R_{ikj}^{\kk\kk k}\ .
\ee
$M_{ij}$ can then be also thought of as encoding
information about the curvature of the transverse space (the second term)
and the spatial part of the energy momentum tensor hidden in what
we call $\tau_{ij}$. 
We can use the supergravity equations of motion to
write $M_{ij}$ explicitely in terms of the field components and derivatives
in the three dimensional transverse space 
\bbb
M_{ij}&=&\frac{1}{2} \lk(\phi_{,i}\phi_{,j}+\phi_{,ij}+R_{ikj}^{\kk\kk k}\re)
-\frac{1}{8} H_{i\mu\nu} H_{j}^{\kk \mu\nu}
-\frac{1}{4} e^{2\phi}
\lk(F^{(2)}_{i\mu} F_{\kk j}^{(2) \mu} -\frac{1}{4} \delta_{ij}
\lk(F^{(2)}\re)^2\re) \nonumber \\
&-&\frac{1}{24} e^{2\phi} \lk(\tilde{F}^{(4)}_{i\alpha\beta\gamma} 
\tilde{F}_{\kk j}^{(4) \alpha\beta\gamma}
-\frac{1}{8} \delta_{ij} 
\lk(\tilde{F}^{(4)}\re)^2\re)-\frac{1}{2} F^{(2)}_{0i,j}\ ,
\eee
with $\tilde{F}^{(4)}\equiv F^{(4)}- C^{(1)}\wedge H^{(3)}$.

If we consider the dynamics of two nearby geodesics at point $P$,
their relative acceleration is related to the separation $z^i$ between them by
the well known equation
\bb
a^i=R_{j0i}^{\kk\kk 0} z^j=-2 M_{ij} z^j\ .
\ee
The condition that the eigenvalues of the matrix $M_{ij}$ must be negative
is simply the statement that the space must be curved
such that two nearby geodesics repel each other. 
The polarization phenomena has to do with the effect of
tidal-like forces in $M_{ij}$ 
expanding the $N$ D0 branes against the binding forces 
due to strings stretched between them. 
There is apparently a scenario where
a balance between these two competing effects is possible and one finds
a non-commutative ellipsoid, encoding data about the conventional gravitational
tidal forces, second derivatives in the dilaton, and the gradient of the
background D0 brane electric field (see equation~\pref{mij}). Note however that
there is an issue of stability, alluded to above,
with regards to this statement.
As presented, these solutions are at the extrema of
the potential, but it is unclear whether they would want to stay there. In particular,
there are directions about the $\Phi^i=0$ point within the Cartan subalgebra of SU(N)
that will drive the energy to arbitrarily small values\footnote{This issue was pointed
out to me by A. Volovich and M. Spradlin.}. These modes may destabilize the ellipsoid as
well. A careful analysis is needed to ascertain this issue. Even if
there exists a stabilization mechanism about the ellipsoidal configuration, the
presence of a dip in the potential energy about $\Phi^i=0$ suggest that these configurations
may not be reached in a generic scenario. We will attempt to elaborate on all of these problems
in an upcoming work~\cite{STABLE}.  For now, looking at our original action, it is easy to read off
the time scale for possible decay. It is set by the typical wavelength of the background
fields. We will see below that this implies that smaller configurations are longer lived. 
It then becomes important to determine the relevant statement
with respect to the time scale useful for observation.

It is also worth emphasizing that the response of the D0 brane configuration
to these tidal forces is somewhat exotic. As depicted in Figure~\ref{fig1}(a),
there are regions of the parameter space where all the $M_i$'s are negative,
yet our solution fails. Too much anisotropy in the
background space is destructive. For example, when $M_1=M_2+M_3$, with
all $M_i$'s negative, the ellipsoid collapses in one direction to become
a disk. This phenomenon is a probe into the internal distributions
of the forces amongst the D0 branes, including interactions resulting
from strings stretched between them.

A good measure of the extent the D0 branes spread out was 
introduced by~\cite{WATISPHERE}.
We denote the size of the configuration by $r^2$ and define
\bb\label{size}
\frac{r^2}{\lambda^2}
\equiv \frac{\Tr\lk\{\Phi^i \Phi^i\re\}}{N}=-\frac{N^2-1}{12} M_{ii}\ .
\ee
Looking back at equation~\pref{maineq}, it is easy to see that the
regime of validity for the expansion in $\lambda$ can be stated 
as the condition
\bb
\lambda^2 \Tr \lk\{ \Phi^i \Phi^j \re\} \del_i \del_j \ll N\ .
\ee
If we write $L$ for the characteristic length scale over which the fields
vary, this implies 
\bb\label{dbicond}
r^2\ll L^2\ .
\ee
The size of the polarized configuration must be much less than the
characteristic length scale over which the background fields vary.
Note that this was also the condition needed so that the trajectory of the center of mass
of the D0 branes is not affected by the details of its polarized shape.
Using equation~\pref{size} in~\pref{dbicond}, we get the relation
\bb\label{arealaw}
N\ll \AA\equiv \frac{L^2}{\alp}\ ,
\ee
where 
$\AA$ is the area constructed using $L$,
in string units\footnote{
We remind the reader that this is a statement in the string frame;
the corresponding equation in the conventional Einstein frame is with
respect to area in Planck units. }. Note that we also need that $L\gg \ls$.
As the bound in equation~\pref{arealaw} is approached, our description
of the problem breaks down. It is a bound 
on the number of degrees of freedom, the D0 branes, that can be
placed at point and consistently be described through the DBI action. More
on this in the Discussion section.

The size of the configuration as given in~\pref{size} is
\bb\label{size2}
r\sim \ls N \lk(\frac{\ls}{L}\re)\ .
\ee
The larger the background length scale $L$, the smaller the size of
the configuration; and the size grows with the number of D0 branes.
This may be viewed as a local manifestation of the UV-IR
relation~\cite{PEETPOLCH}; simply
put, the ``dispersion relation'' of our probes in the curved background.
And for $N<L/\ls$, where $L/\ls$ is
always much greater than one, the size of the configuration is substringy,
yet we are in a valid regime for the computation. As is well known,
the D0 branes are naturally good probes of Planck scale distances in
space.

The physical data encoded in $M_{ij}$ consisted of three numbers,
and a Lie algebraic structure is enough to resolve this information,
transfering it into the shape of the non-commutative ellipsoid. Physically,
the matrix $M_{ij}$ gets encoded in a selective set of stretched
open strings.
We will see in Section~\ref{qgsec} that, to resolve more structure
of the background space, such as higher derivatives of the metric, more
``links'' between the D0 brane would be needed, and a Lie algebraic
structure is not enough.

\subsection{A non-commutative sphere and balance of forces}
\label{balancesec}

In this section,
we consider solutions to~\pref{mateq} with both 
$\tilde{N}$ and $M_{ij}$ non-zero. Solving the full
anisotropic case involves as much pleasure as solving
Maxwell's equations in a cavity of arbitrary shape, void of any symmetries.
Most of the dynamics of the competition between the various forces
can be demonstrated by considering an isotropic configuration, where
the eigenvalues of the matrix $M_{ij}$ are all equal
$M_1=M_2=M_3\equiv M$; consequently, the solution is a non-commutative
sphere~\cite{WHN,WATISPHERE} 
and we write the eigenvalues of $g_{ij}$ as $g_1=g_2=g_3\equiv g$.
We then have the equation
\bb
g^2- 3 \tilde{N} g +M=0\ ,
\ee
with solutions
\bb\label{gpm}
g_{\pm}=\frac{3\tilde{N}}{2} \lk(1\pm\lk(1-\frac{4 M}{9\tilde{N}^2}\re)^{1/2}
\re)\ ,
\ee
with the condition
\bb\label{mineqone}
M\leq \frac{9}{4} \tilde{N}^2\ .
\ee
We learned from our previous discussion in Section~\ref{ellipsoidsec} that
the effect of $\tilde{N}$
is to create repulsive forces between the D0 branes. This means
we may now expect solutions with 
$M>0$ as well, where equation~\pref{mineqone}
becomes relevant.

To decide between the two solutions in~\pref{gpm}, we evaluate the
potential energy of the configuration
\bb
V=\frac{3}{\gs\ls} \lambda^2 \frac{N(N^2-1)}{24} g_\pm^3 
\lk(2\tilde{N}-g_\pm\re)\ .
\ee
And requiring $V\leq 0$, we arrive at the condition
\bb
\lk(1 \pm \lk(1-\frac{4 M}{9\tilde{N}^2}\re)^{1/2}\re)
\lk(1\mp 3 \lk(1-\frac{4 M}{9\tilde{N}^2}\re)^{1/2}\re) \leq 0\ .
\ee
We then can easily determine the following possibilities:
for $M>0$, the only solution is given by $g_+$, subject to the condition
\bb
M<2\tilde{N}^2\ ,
\ee
which is stronger than~\pref{mineqone}, and hence prevails. For $M<0$,
equation~\pref{mineqone} is satisfied, and we have both solutions $g_\pm$
being possible. The one that prevails is the one with lower energy.
Looking at~\pref{gpm}, we see that we have $|g_+|>|g_-|$. Hence,
if
\bb
g_+ \lk(2\tilde{N}-g_+\re)<g_-\lk(2\tilde{N}-g_-\re) \Rightarrow V_+<V_-\ .
\ee
Rearranging this equation, and using~\pref{gpm}, we find that the
energetically favored solution for $M<0$ is $g_+$ again.
We arrive at the simple phase structure shown in Figure~\ref{fig1}(b).
In the shaded region, our solutions are disfavored even classically.
We see that the
interactions appearing in~\pref{maineq} 
may predict interesting phase structures
of vacuum configurations of D brane probes; one that results from the
competition of the various couplings to the background fields. 

\section{Beyond Lie algebras}
\label{qgsec}

It is apparent from our discussion
in the previous sections that the Lie algebraic structure was exhausted
in the process of
encoding the space-time data into the D0 brane matrices.
As more details and moments get probed by higher order couplings,
it is then natural to ask how does this additional information get stored
in the $\Phi^i$ matrices. 
In this section, we explore terms in~\pref{maineq} 
cubic in $\lambda$, by
outlining a formal recursive
prescription, involving an expansion in $\lambda$,
which allows us to go beyond the Lie algebraic structure.
We again confine our discussion to a three dimensional
subspace of the transverse space, even though many of the equations
we will write are relevant to more general cases as well. 

The equations of motion we consider
are a specialization of~\pref{eomthree}
\bb\label{cubiceq}
2 \Phi^i M_{ni} +\C{\Phi^i}{\C{\Phi^i}{\Phi^n}}
+\frac{3}{2} \lambda \AC{\Phi^i}{\Phi^j}T_{nij}=0\ ,
\ee
with $M_{ij}$ and $T_{ijk}$ totally symmetric tensors.
From the structure of~\pref{cubiceq}, we see that the Lie algebra
\bb
\C{\Phi^i}{\Phi^j}=C^{ij}_{\kk k} \Phi^k
\ee
would account for the term involving $M_{ij}$ as described in the
previous sections. But then, the term involving the anti-commutator
would introduce matrices outside the space spanned by the
original $\Phi^i$'s; we have
\bb\label{anticom}
\AC{\Phi^i}{\Phi^j}= 2 h^{ij} {\bf 1} + \sigma^{ij}\ ,
\ee
where
\bb
h^{ij}\equiv \Tr\lk\{\Phi^i \Phi^j \re\}\ .
\ee
Generically, the six $N\times N$ matrices $\sigma^{ij}$ 
are not in the SU(2) algebra. Note however that,
in the case $N=2$, the Pauli matrices anticommute such that the
$\sigma^{ij}$'s are zero. These matrices arise as we add more and more 
D0 branes, increasing the size of the matrices. Physically, this is an
interesting statement. With a few D0 branes, we can resolve so
much of the spacetime structure. As we introduce higher derivative
perturbations in the background fields, polarized
configurations with fewer D0 branes will tend to become unstable sooner
than configurations with larger number of D0 branes. This is because,
to resolve additional structure in the background fields, one adds
more stretched strings or ``links'' between the D0 branes; and this
pool of matrix data is limited by the size of the matrices, the $N^2$
entries in the $\Phi^i$'s. This is a generic idea; an identical
phenomenon can be visualized for example in electromagnetism with
respect to the competition between the number of charges and higher moments
of electric field backgrounds. We assume that $N$ is large enough so that
a solution to the problem at hand is possible. A naive counting 
suggest a number greater than $4$.
The idea involved in solving~\pref{cubiceq} is
to ``add additional links'' between the D0 branes
by turning on off-diagonal elements in the D0 brane matrices, so 
as to balance the shearing forces due to the background field $T_{nij}$.

Consider a solution of the form
\bb\label{ansatz}
\Psi^i = \Phi^i + \lambda e^i_{kl} \sigma^{kl}+\lambda \gamma^i {\bf 1}
+O(\lambda^2)\ .
\ee
The $\Phi^i$'s satisfy the SU(2) algebra. To order $\lambda$, we are perturbing
by the traceless hermitian matrices $\sigma^{ij}$
appearing on the right side of~\pref{anticom}.
These contain the additional
``links'' that need to be activated between the D0 branes. The c-number
parameters $e^i_{kl}$ play the role of vielbeins; they connect
the space time index to the matrix space spanned by the $\sigma^{ij}$'s.
Looking at equation~\pref{cubiceq},
we see that adding the $\sigma^{ij}$'s is not enough due to the
term proportional to the identity in~\pref{anticom}. This requires adding
the term involving $\gamma^i$ in~\pref{ansatz}. Physically, this
is the statement that some of the higher moments in $T_{nij}$ shift
the center of mass of the configuration; these are the modes
given by $T_{nij} h^{ij}$. The presence of the identity in~\pref{ansatz} is
needed to account for effects of the size of the configuration on the trajectory
of the center of mass. The U(1) and SU(N) in U(N) do not decouple at this order,
and this is incorporated in the trace part of $T_{nij}$. The resulting configuration
is shifted off-center.

We now can easily compute 
\bb\label{closure}
\C{\Phi^k}{\sigma^{ij}}=C^{ki}_{\kk l} \sigma^{lj}+C^{kj}_{\kk l} \sigma^{il}\ .
\ee
where we have used the fact that
\bb
C^{ijk}\sim C^{ij}_{\kk l} h^{lk}
\ee
is total antisymmetric on $i,j,k$, given that $h^{ij}$ is propotional to
the Cartan-Killing metric of SU(2). Equation~\pref{closure} is an important
ingredient of our prescription. It is related to a well known identity
and corresponds to the statement that the commutator of $\Phi^i$ with
the $\sigma^{kl}$'s closes onto the space spanned by the $\sigma^{kl}$'s.
We compute the commutator of our ansatz~\pref{ansatz}
\bbb\label{genalgebra}
\C{\Psi^i}{\Psi^n}&=&C^{in}_{\kk m} \Phi^m
+2 \lambda \lk( e^n_{kl} C^{il}_{\kk m}
-e^i_{kl} C^{nl}_{\kk m}\re) \sigma^{km} +O(\lambda^2) \nonumber \\
&=&C^{in}_{\kk k} \Psi^k + 2\lambda D^{in}_{ml} \sigma^{ml}
-{\bf 1}\lambda C^{in}_{\kk k} \gamma^k + O(\lambda^2)\nonumber \\
&=&C^{in}_{\kk k} \Psi^k
-\lambda C^{in}_{\kk k} \gamma^k {\bf 1}
- 4\lambda D^{in}_{ml} h^{ml} {\bf 1}
+ 2\lambda D^{in}_{ml} \AC{\Psi^m}{\Psi^l}+ O(\lambda^2)\ ,
\eee
where we have introduced
\bb\label{Dterms}
D^{ij}_{ml}\equiv e^j_{kl} C^{ik}_{\kk m} - e^i_{kl} C^{jk}_{\kk m}
-\frac{1}{2} e^k_{lm} C^{ij}_{\kk k}\ .
\ee
These D-structure constants are symmetric in the lower indices,
and antisymmetric in the upper.
We substitute all these relations into~\pref{cubiceq} and 
obtain the following equations:

\begin{itemize}
\item To $O(\lambda)$, we have terms involing only the $\Phi^i$
as before. We get equations that determine the structure constants
$C^{ij}_{\kk k}$ in terms of $M_{ij}$
\bb\label{mnm}
2 M_{nm} + C^{in}_{\kk l} C^{il}_{\kk m}=0\ .
\ee

\item To $O(\lambda^2)$, the trace part~\pref{cubiceq} determines the 
$\gamma^i$'s
\bb
3 T_{nij} h^{ij} +2 M_{ni} \gamma^i=0\ .
\ee

\item To $O(\lambda^2)$, we have terms involving
the $\sigma^{ij}$'s; and we find equations that determine the $D_{ij}^{kl}$
in terms of $T_{nij}$
\bb\label{Deq}
\frac{3}{4} T_{nlj} + C^{in}_{\kk k} D^{ik}_{lj}+ 2 C^{im}_{\kk j} D^{in}_{lm}
=0\ .
\ee
In deriving this equation, we have made use of equation~\pref{mnm}
that appears at a lower order in $\lambda$. Note that the vielbeins
do not appear explicitely, and we end up solving for the D-structure
constants.
Formally, from these, we can detemine the veilbeins $e^{kl}_i$. Naively,
it also appears we have more unknowns than equations, and there is an issue
about the uniqueness of solutions to~\pref{Deq}. Such freedom may correspond
to symmetries, or some of this information may be needed when other
fields get turned on.
\end{itemize}

The generalized algebra in~\pref{genalgebra} is an extension of the Lie
algebra defined by the $O(\lambda^2)$ solution. It has structure similar
to a q-deformed algebra~\cite{QWESS,MSSW}. 
It is instructive
to briefly 
comment on how this structure is realized explicitely in matrix space.
The SU(2) part of the algebra in an $N$ dimensional representation
correspond to matrices of the form, schematically
\bb
\lk(
\begin{array}{cccccc}
x & x & 0 & 0 & 0 & \\
x & x & x & 0 & 0 & \\
0 & x & x & x & 0 & \cdots \\
0 & 0 & x & x & x & \\
0 & 0 & 0 & x & x & \\
  &   & \vdots &  &  &
\end{array}
\re)
\ee
The corresponding $\sigma^{ij}$'s by which we perturb the solution 
span matrices of the form
\bb
\lambda
\lk(
\begin{array}{cccccc}
x & x & x & 0 & 0 & \\
x & x & x & x & 0 & \\
x & x & x & x & x & \cdots \\
0 & x & x & x & x & \\
0 & 0 & x & x & x & \\
  &   & \vdots &  &  &
\end{array}
\re)
\ee
We see that the additional data is encoded in ``next to nearest neighbour''
links between the D0 branes. Perhaps there is a general pattern where
we explore the matrix space starting
along the diagonal and moving outward as we expand to higher
orders in $\lambda$.

\section{A non-compact solution}
\label{ncsec}

In this section, we come back to~\pref{eomtwo}, 
with the field $A_i$ set to zero 
\bb\label{basic}
2 \Phi^i M_i = \C{\Phi^i}{\C{\Phi^n}{\Phi^i}}\ ,
\ee
looking for solutions beyond the pyramid defined by Figure~\ref{fig1}(a). 
We consider
the case where one eigenvalue of $M_{ij}$ is zero, and the other two
{\em positive} and equal. It appears from our previous
discussion that a compact configuration is not possible.
We will therefore look for non-compact solutions, such as the
one that solves the flat space case $\C{\Phi^1}{\Phi^2}=i\theta {\bf 1}$
considered in~\cite{MAT1,SEIBBACK}. These vacua will necessarily have infinite
energy but may be stabilized by appropriate boundary conditions at
infinity. We may also want to consider the Matrix theory limit
$\gs,\ls\rightarrow 0$ with $\ls/\gs^{1/3}$ fixed, so that terms
of order $\lambda^3$ and beyond in the action can be
ignored all the way to asymptotic infinity in spacetime.

Working in a diagonal basis of the $M_{ij}$ matrix, 
the matrix entries are chosen as
\bb
M_1=0\ \ \ ,\ \ \ 
M_2=M_3=M>0\ .
\ee
Inspecting equation~\pref{basic}, we can easily write 
the solution as the closed algebra
\bb
\C{\Phi^2}{\Phi^3}= i\theta {\bf 1}\ \ \ ,\ \ \
\C{\Phi^1}{\Phi^3}= - i \sqrt{M} \Phi^2\ \ \ ,\ \ \ 
\C{\Phi^1}{\Phi^2}= - i \sqrt{M} \Phi^3\ ;
\ee
and we can represent $\Phi^1$ as
\bb
\Phi^1=\frac{\sqrt{M}}{2\theta} \lk ( (\Phi^3)^2-(\Phi^2)^2 \re)\ ,
\ee
This seems to describe a non-commutative hyperboloid;
a microscopic description of a deformed D2 brane extending all the way
to infinity. Note that,
to realize this algebra, we need to take matrices
of infinite size $N\rightarrow \infty$.

The potential energy of this configuration is given by
\bb
V=\frac{\lambda^2}{\gs \ls}
\lk (M \Tr \lk \{ (\Phi^2)^2+(\Phi^3)^2 \re \} +\frac{\theta^2}{2} N \re)\ .
\ee
The second term is the same 
for the plane configuration 
in flat space as well~\cite{MAT1,SEIBBACK} and we may choose
to substract it to quantify the energy content of our configuration.
We can evaluate the trace over the Heisenberg operators by 
``regulating'' the divergence with $N$
\bb
\Tr q^2=\frac{\theta}{2} \sum_{j=0}^{N-1}\ 2j+1=\frac{\theta}{2} N^2\ .
\ee
We then consider the excess energy
\bb
\frac{V}{N^2}\rightarrow\frac{3}{4} \frac{1}{\gs\ls} \lk(\lambda \theta\re)
\lk(\lambda M\re)\sim \frac{1}{\gs\ls} \frac{\Theta}{(L/\ls)^2}\ .
\ee
$\Theta$ is the scale of non-commutativity on the hyperboloid in string
units. This excess energy seems to be distributed amongst all $N^2$
entries of the matrices.

In general, we may expect exotic solutions such as this in regions of
the background field parameter space where compact configurations cannot
exists. We may expect that we are
allowed to use contractions and deformations of SU(2),
as long as we end up with a consistant, closed algebras that
is energetically or dynamically favoured.

\section{Discussion}
\label{discussion}

Two statements from our analysis of the non-commutative
ellipsoid are worth further elaboration. Equation~\pref{arealaw}
determines the regime of validity of the DBI expansion. It places
a bound on the number of D0 branes in terms of the
length scale over which the background fields are varying. It is
a statement about limiting one bit of information per Planck area,
with the measure of area being defined 
in a local manner, using the
characteristic scales of the background\footnote{
Presumably, the statement is sensitive to the fact that we
restricted the dynamics to a three dimensional subset of the transverse
space and that the resulting object is a D2 brane; hence the quadratic power
in $L$. 
It is an interesting problem to
understand the same issue in higher dimensions.
}.
As we tune the wavelength $L$ of background fields to smaller values,
the configuration of D0 branes expands in size, until matching the
characteristic size of the background field variations. At this point,
$L$ is small enough that we have one D0 per Planck area. 
Our analysis is about D0 branes acting as probes to
the structure of spacetime, ignoring 
back reaction effects due to the D0 branes themselves.
As this bound gets saturated, 
our formalism breaks down, and the situation needs
to be described within the context of the full string theory.
We may expect that back reaction effects will become important before reaching
the critical point, and equation~\pref{arealaw} should be interpreted
as an approximate scaling relation.
Note however that, at the saturation point, background curvature scales
can be very small for large values of $N$, well within the supergravity
approximation regime.
In view of independent observations about Holographic bounds and
black hole entropy, we may interpret~\pref{arealaw} as more that just
a statement restricting 
the regime of validity of an approximation scheme; but one that is
rooted in fundamental physics.

The second interesting point has to do with a local formulation of
the UV-IR correspondence. Equation~\pref{size2} shows that the 
D0 brane ellipsoid shrinks in size with larger wavelengths of the
background space. Let us imagine perturbing the vacuum 
ellipsoidal configuration so as
to write the theory of surface fluctuations as was done 
in~\cite{AOKI,SEIBBACK,VIBDAS}.
We can represent the matrix algebra over the space of smooth functions
by introducing the appriopriate star product~\cite{SWNC,AOKI,SEIBBACK,LI}. 
We would be
describing the microscopic dynamics of a D2 brane-like object 
contructed from D0 branes. The 2+1 dimensional worldvolume theory 
is non-commutative and lives on a compact space.
The size of the configuration is given by~\pref{size2}; we remind the
reader that the metric of relevance is~\pref{inertial}, \ie\ it is flat.
Hence, the IR cutoff in this theory is
\bb\label{ir}
\Sigma\sim r\sim \ls \frac{N}{L/\ls}\ .
\ee
In~\cite{AOKI,SEIBBACK}, an interesting general statement was made
relating the IR cutoff $\Sigma$, the non-commutativity parameter
$\theta$ and the number of D0 branes that underly a non-commutative
field theory
\bb\label{uv}
N\sim\frac{\Sigma}{\sqrt{\theta}}\ .
\ee
The non-commutativity scale $\theta$ plays the role
of UV regulator; equation~\pref{uv} states that the number of D0
branes underlying the non-commutative dynamics is
given by the ratio of the IR to the UV cutoff; this is a simple yet important
statement of a general character. We can use it to estimate
the scale of non-commutativity in the worldvolume theory resulting
from perturbing the ellipsoidal configuration. Using~\pref{ir} as
the IR cutoff, we find (in the string frame)
\bb
\sqrt{\theta}\sim \frac{\alp}{L}\ .
\ee
The larger the wavelengths in the background space, the smaller the
length scale of non-commutativity in the world-volume theory. This is a 
local manifestation of the usual UV-IR correspondence
$U=r/\alp$~\cite{MALDA,PEETPOLCH}.

Restricting the discussion to a three dimensional subspace led to
configurations which are microscopic 
realizations of D2 branes~\cite{HASHITZMOR,EPRABIN}.
One should take note of the fact that these polarized states resulted 
from tidal-like forces, without the need to turn on the 
RR gauge fields corresponding to a higher dimensional brane. Yet, the 
end result will carry D2 brane charge as dictated by the couplings
appearing in~\pref{maineq}. Different
parts of the droplets of D0 branes ``fall'' with different accelerations; 
hence the D0 branes expand while retaining a coherent shape. 
This effect is achieved by gravitational
tidal forces, gradients in the background D0 brane one form gauge field,
and non-zero second derivatives of the dilaton field.

The stability of the configurations we found is an issue that needs to 
be determined. This can be done by perturbing the solutions and looking for
tachyonic modes.  And in case of decay, the time scale is important to determine the
physical relevance of these configurations. As presented, one should view
our solutions as extrema of the action with interesting properties that
probe the issue on how to distribute smooth spacetime information into
matrices. We will tackle the stability problem in detail in~\cite{STABLE}.

On a few technical notes, it is worthwhile pointing out that, in
expanding~\pref{maineq}, some simplifications arose due to the
symmetrized trace prescription introduced in~\cite{TSEYTDBI}. It is known
that this procedure fails at higher orders~\cite{BAIN}, 
but none of our calculations
probed this regime. Another aspect has to do with 
representing the non-commutative algebras of matrices 
on the space of functions. 
It may be intructive to study the worldvolume theories on the polarized
configurations, as a prelude to connecting to a smooth macroscopic
D2 brane picture. The star operator will arise in this context, and
the mathematical foundations of this, for both Lie and q-deformed algebras,
have recently been explored in~\cite{QWESS,MSSW,KLM,VACARIU}. 
In particular, there is 
a well-defined star product associated with the D-structure constants
introduced in~\pref{Dterms}.

There are several immediate extensions of these ideas that are of
interest. For one, it would be useful to understand similar polarization
phenomena in higher than three dimensions. We should expect for example
D4 or D6 branes arising from the polarizing effects of tidal forces.
The algebras of relevance maybe SU(2)$\times$SU(2) and SU(3). The
eight generators of the latter minus its two Cassimirs perhaps provide
the appropriate embedding of the worlvolume in the spacetime. Given
the subtleties associated with the five brane, such an approach may be
too naive after all~\cite{MEYERS}.

It would be interesting to understand the pattern of higher order
effects in the DBI action in the structure of vacuum solutions. 
The question is about the forms of generalized algebras that can arise
in encoding information about spacetime fields into matrices
as an expansion in the moments of the fields. It would also be helpful
to write realizations of these generalized algebras in explicit
examples, to develop intuition about the dynamics involved.

In~\cite{MINAS}, the Chern-Simmons term was generalized to include
the effects of additional couplings to the spacetime curvature. The
effects of these
need to be considered in a consistent analysis
specially when background curvature scales are a large. They
would introduce couplings with even powers of the Ricci tensor and
the RR gauge fields.
Another interesting issue is to consider dynamical situations; such as
when a gravitational wave sweeps past a configuration of D0 branes. 
This seting will explore a new set of interaction terms arising
in~\pref{maineq} which we did not consider.

Finally, an important issue may be to understand this polarization
phenomena in the presence of 
time-space non-commutativity~\cite{GMSS,KLEBMALDA,YONEYA}. 
The relevant setup may be to study D-instanton dynamics~\cite{IKKT} 
near black hole
horizons. The flipping of the light-cone at a horizon may necessitate
consideration of non-commutation of time and space. 
We hope to report on this issue in the future.

\section{Appendix: A few technical details}

In this appendix, we sketch some of the computational
details of the main text.
The main ingredients in expanding~\pref{maineq} to order $\lambda^3$
under the conditions outlined in Section~\ref{setupsec} are 
the expansion of the supergravity field as in~\pref{expansion},
and the expansion of $\det\ Q$. The term containing $Q^{-1}-\delta$
of~\pref{maineq} never arises in our discussion, and the pull-back
is trivial in the static scenario.
The determinant of the $Q$ matrix expands to
\bbb
\lk(\det Q\re)^{1/2}&=&
1+i\frac{\lambda^2}{2} \C{\Phi^i}{\Phi^k} \Phi^l B_{ki,l}
-\frac{\lambda^2}{4} \C{\Phi^i}{\Phi^j}\C{\Phi^i}{\Phi^j} \nonumber \\
&+&i\frac{\lambda^3}{4} \C{\Phi^i}{\Phi^k} \Phi^m \Phi^l B_{ki,ml}
+\frac{\lambda^3}{2} \C{\Phi^i}{\Phi^j}\C{\Phi^j}{\Phi^m} \Phi^l B_{mi,l}
\nonumber \\
&-&i \frac{\lambda^3}{6} \C{\Phi^i}{\Phi^k}\C{\Phi^k}{\Phi^l}\C{\Phi^l}{\Phi^i}
+O(\lambda^4) \ .
\eee
The last two terms do not contribute; the first of these vanishes due
to the symmetry in the $i$ and $m$ indices using the symmetrized
trace prescription; the second vanishes also because it arises in the 
symmetrized trace.

The Chern-Simmons terms expand to
\bbb
S_{CS}&=&\frac{1}{\gs \ls}\int \Str \lk\{
P\lk[
C^{(1)}+ i \lambda i_\Phi i_\Phi C^{(3)}
+i \lambda^2 i_\Phi i_\Phi \lk( C^{(1)} B,i\re) \Phi^i \re. \re. \nonumber \\
&-&
\lk. \lk.  \frac{\lambda^2}{2} \lk( i_\Phi i_\Phi \re)^2 C^{(5)} + O(\lambda^3)
\re]
\re\}\ .
\eee
with
\bb
P\lk[C^{(1)}\re]=C_0^{(1)} + \lambda \Phi^i C^{(1)}_{0,i}
+\frac{\lambda^2}{4} \AC{\Phi^i}{\Phi^j} C^{(1)}_{0,ij} + O(\lambda^3)\ ;
\ee
\bb
i\lambda P\lk[ i_\Phi i_\Phi \lk(C^{(3)}+ C^{(1)} B\re) \re]=
i\frac{\lambda}{2} \C{\Phi^j}{\Phi^i} \lk(
C^{(3)}_{ij0} + \lambda \Phi^k C^{(3)}_{ij0,k}\re)
+i\frac{\lambda^2}{2} \C{\Phi^j}{\Phi^i} \Phi^k C^{(1)}_0 B_{ij,k}\ ;
\ee
\bb
-\frac{\lambda^2}{2} P\lk [\lk(i_\Phi i_\Phi \re)^2 C^{(5)}\re]=
-\frac{\lambda^2}{8} \C{\Phi^l}{\Phi^k}\C{\Phi^j}{\Phi^i} C^{(5)}_{ijkl0}
\ .
\ee

Puting everything together, we get the actions given in~\pref{actiontwo} 
and~\pref{actionthree}.

\paragraph{\bf Acknowledgments:}
I thank P. Argyres, T. Becher, M. Spradlin, H. Tye, and A. Volovich for discussions.
This work was supported by NSF grant 9513717.


\begin{thebibliography}{10}

\bibitem{MEYERS}
R.~C. Myers, ``Dielectric-branes,'' {\em JHEP} {\bf 12} (1999) 022,
  \href{http://xxx.lanl.gov/abs/hep-th/9910053}{{\tt hep-th/9910053}}.

\bibitem{WATIDBI}
I.~Washington~Taylor and M.~V. Raamsdonk, ``Multiple D0-branes in weakly curved
  backgrounds,'' {\em Nucl. Phys.} {\bf B558} (1999) 63--95,
  \href{http://xxx.lanl.gov/abs/hep-th/9904095}{{\tt hep-th/9904095}}.

\bibitem{GIANT}
J.~McGreevy, L.~Susskind, and N.~Toumbas, ``Invasion of the giant gravitons
  from Anti-de Sitter space,'' {\em JHEP} {\bf 06} (2000) 008,
  \href{http://xxx.lanl.gov/abs/hep-th/0003075}{{\tt hep-th/0003075}}.

\bibitem{HASHHIRITZ}
A.~Hashimoto, S.~Hirano, and N.~Itzhaki, ``Large branes in AdS and their field
  theory dual,'' {\em JHEP} {\bf 08} (2000) 051,
  \href{http://xxx.lanl.gov/abs/hep-th/0008016}{{\tt hep-th/0008016}}.

\bibitem{DASGIANT}
S.~R. Das, A.~Jevicki, and S.~D. Mathur, ``Giant gravitons, BPS bounds and
  noncommutativity,'' \href{http://xxx.lanl.gov/abs/hep-th/0008088}{{\tt
  hep-th/0008088}}.

\bibitem{DTVPOL}
S.~R. Das, S.~P. Trivedi, and S.~Vaidya, ``Magnetic moments of branes and giant
  gravitons,'' \href{http://xxx.lanl.gov/abs/hep-th/0008203}{{\tt
  hep-th/0008203}}.

\bibitem{SWNC}
N.~Seiberg and E.~Witten, ``String theory and noncommutative geometry,'' {\em
  JHEP} {\bf 09} (1999) 032, \href{http://xxx.lanl.gov/abs/hep-th/9908142}{{\tt
  hep-th/9908142}}.

\bibitem{AOKI}
H.~Aoki {\em et.~al.}, ``Noncommutative Yang-Mills in IIB Matrix model,'' {\em
  Nucl. Phys.} {\bf B565} (2000) 176--192,
  \href{http://xxx.lanl.gov/abs/hep-th/9908141}{{\tt hep-th/9908141}}.

\bibitem{SEIBBACK}
N.~Seiberg, ``A note on background independence in noncommutative gauge
  theories, matrix model and tachyon condensation,'' {\em JHEP} {\bf 09} (2000)
  003, \href{http://xxx.lanl.gov/abs/hep-th/0008013}{{\tt hep-th/0008013}}.

\bibitem{DKPS}
M.~R. Douglas, D.~Kabat, P.~Pouliot, and S.~H. Shenker, ``D-branes and short
  distances in string theory,'' {\em Nucl. Phys.} {\bf B485} (1997) 85--127,
  \href{http://xxx.lanl.gov/abs/hep-th/9608024}{{\tt hep-th/9608024}}.

\bibitem{POLCHV1}
J.Polchinski, {\em String Theory, Vol. 1}.
\newblock Cambridge University Press, 1998.

\bibitem{PEETPOLCH}
A.~W. Peet and J.~Polchinski, ``UV / IR relations in AdS dynamics,''
  \href{http://xxx.lanl.gov/abs/hep-th/9809022}{{\tt hep-th/9809022}}.

\bibitem{QWESS}
J.~Wess, ``q-deformed Heisenberg algebras,''
  \href{http://xxx.lanl.gov/abs/math-ph/9910013}{{\tt math-ph/9910013}}.

\bibitem{MSSW}
J.~Madore, S.~Schraml, P.~Schupp, and J.~Wess, ``Gauge theory on noncommutative
  spaces,'' {\em Eur. Phys. J.} {\bf C16} (2000) 161,
  \href{http://xxx.lanl.gov/abs/hep-th/0001203}{{\tt hep-th/0001203}}.

\bibitem{MILLARWATI}
K.~Millar, W.~Taylor, and M.~V. Raamsdonk, ``D-particle polarizations with
  multipole moments of higher- dimensional branes,''
  \href{http://xxx.lanl.gov/abs/hep-th/0007157}{{\tt hep-th/0007157}}.

\bibitem{HO}
P.-M. Ho, ``Fuzzy sphere from Matrix model,''
  \href{http://xxx.lanl.gov/abs/hep-th/0010165}{{\tt hep-th/0010165}}.

\bibitem{GARMEY1}
M.~R. Garousi and R.~C. Myers, ``World-volume interactions on D-branes,'' {\em
  Nucl. Phys.} {\bf B542} (1999) 73--88,
  \href{http://xxx.lanl.gov/abs/hep-th/9809100}{{\tt hep-th/9809100}}.

\bibitem{GARMEY2}
M.~R. Garousi and R.~C. Myers, ``World-volume potentials on D-branes,''
  \href{http://xxx.lanl.gov/abs/hep-th/0010122}{{\tt hep-th/0010122}}.

\bibitem{TSEYTDBI}
A.~A. Tseytlin, ``On non-abelian generalisation of the Born-Infeld action in
  string theory,'' {\em Nucl. Phys.} {\bf B501} (1997) 41--52,
  \href{http://xxx.lanl.gov/abs/hep-th/9701125}{{\tt hep-th/9701125}}.

\bibitem{MISNER}
C.~Misner, K.~Thorne, J.~Wheeler, ``Gravitation,'' W.~H.~Freeman and company,
New York.

\bibitem{WATISPHERE}
D.~Kabat and I.~Washington~Taylor, ``Spherical membranes in Matrix theory,''
  {\em Adv. Theor. Math. Phys.} {\bf 2} (1998) 181,
  \href{http://xxx.lanl.gov/abs/hep-th/9711078}{{\tt hep-th/9711078}}.

\bibitem{WHN}
B.~de~Wit, J.~Hoppe, and H.~Nicolai, ``On the quantum mechanics of
  supermembranes,'' {\em Nucl. Phys.} {\bf B305} (1988) 545.

\bibitem{MAT1}
T.~Banks, W.~Fischler, S.~H. Shenker, and L.~Susskind, ``M theory as a Matrix
  model: A conjecture,'' {\em Phys. Rev.} {\bf D55} (1997) 5112--5128,
  \href{http://xxx.lanl.gov/abs/hep-th/9610043}{{\tt hep-th/9610043}}.

\bibitem{LI}
M.~Li, ``Strings from IIB matrices,''
{\em Nucl. Phys.} {\bf B499}, (1997) 149,
\href{http://xxx.lanl.gov/abs/hep-th/9612222}{{\tt hep-th/9612222}}.

\bibitem{VIBDAS}
S.~R. Das, A.~Jevicki, and S.~D. Mathur, ``Vibration modes of giant
  gravitons,'' \href{http://xxx.lanl.gov/abs/hep-th/0009019}{{\tt
  hep-th/0009019}}.

\bibitem{MALDA}
J. Maldacena, ``The Large N Limit of Superconformal Field Theories 
  and Supergravity '', \href{http://xxx.lanl.gov/abs/hep-th/9711200}{{\tt
  hep-th/9711200}}.

\bibitem{HASHITZMOR}
A.~Hashimoto and N.~Itzhaki, ``On the hierarchy between non-commutative and
  ordinary supersymmetric Yang-Mills,'' {\em JHEP} {\bf 12} (1999) 007,
  \href{http://xxx.lanl.gov/abs/hep-th/9911057}{{\tt hep-th/9911057}}.

\bibitem{EPRABIN}
S.~Elitzur, B.~Pioline, and E.~Rabinovici, ``On the short-distance structure of
  irrational non- commutative gauge theories,'' {\em JHEP} {\bf 10} (2000) 011,
  \href{http://xxx.lanl.gov/abs/hep-th/0009009}{{\tt hep-th/0009009}}.

\bibitem{KLM}
P.~Kosinski, J.~Lukierski, and P.~Maslanka, ``Local field theory on
  kappa-Minkowski space, star products and noncommutative translations,''
  \href{http://xxx.lanl.gov/abs/hep-th/0009120}{{\tt hep-th/0009120}}.

\bibitem{BAIN}
P. Bain, ``On the non-abelian Born-Infeld action,''
  \href{http://xxx.lanl.gov/abs/hep-th/9909154}{{\tt
  hep-th/9909154}}.

\bibitem{VACARIU}
S.~I. Vacaru, ``Gauge and Einstein gravity from non-abelian gauge models on
  noncommutative spaces,'' \href{http://xxx.lanl.gov/abs/hep-th/0009163}{{\tt
  hep-th/0009163}}.

\bibitem{MINAS}
S.~F. Hassan and R.~Minasian, ``D-brane couplings, RR fields and Clifford
  multiplication,'' \href{http://xxx.lanl.gov/abs/hep-th/0008149}{{\tt
  hep-th/0008149}}.

\bibitem{GMSS}
R.~Gopakumar, S.~Minwalla, N.~Seiberg, and A.~Strominger, ``OM theory in
  diverse dimensions,'' \href{http://xxx.lanl.gov/abs/hep-th/0006062}{{\tt
  hep-th/0006062}}.

\bibitem{KLEBMALDA}
I.~R. Klebanov and J.~Maldacena, ``1+1 dimensional NCOS and its U(N) gauge
  theory dual,'' \href{http://xxx.lanl.gov/abs/hep-th/0006085}{{\tt
  hep-th/0006085}}.

\bibitem{YONEYA}
T.~Yoneya, ``String theory and space-time uncertainty principle,''
  \href{http://xxx.lanl.gov/abs/hep-th/0004074}{{\tt hep-th/0004074}}.

\bibitem{IKKT}
N.~Ishibashi, H.~Kawai, Y.~Kitazawa, and A.~Tsuchiya, ``A large-N reduced model
  as superstring,'' {\em Nucl. Phys.} {\bf B498} (1997) 467--491,
  \href{http://xxx.lanl.gov/abs/hep-th/9612115}{{\tt hep-th/9612115}}.

\bibitem{STABLE}
To appear.

\end{thebibliography}

\providecommand{\href}[2]{#2}\begingroup\raggedright\endgroup

\end{document}